\documentclass[12pt]{article}

\usepackage{amssymb}
\usepackage[dvips]{graphicx}
\usepackage[dvips]{psfrag}
\usepackage{wick}
\usepackage{cite}

\newcommand {\beq}{\begin{equation}}
\newcommand {\eeq}{\end{equation}}
\newcommand {\beqa}{\begin{eqnarray}}
\newcommand {\eeqa}{\end{eqnarray}}
\newcommand {\n}{\nonumber \\}
\newcommand{\tr}{\mbox{tr}}
\newcommand {\Tr}{\mbox{Tr}}

\renewcommand{\theequation}{\thesection.\arabic{equation}}

\begin{document}
\setlength{\oddsidemargin}{0cm}
\setlength{\baselineskip}{7mm}

\begin{titlepage}
\begin{normalsize}
\begin{flushright}
\begin{tabular}{l}
March 2006
\end{tabular}
\end{flushright}
  \end{normalsize}

~~\\

\vspace*{0cm}
    \begin{Large}
       \begin{center}
         {World-Sheets from $\mathcal{N}$ = 4 Super Yang-Mills}
       \end{center}
    \end{Large}
\vspace{1cm}

\begin{center}
           Matsuo S{\sc ato}\footnote
            {
e-mail address : 
sato@pas.rochester.edu}\\
      \vspace{1cm}
       
         {\it Department of Physics and Astronomy}\\
            {\it University of Rochester}\\
               {\it Rochester, NY 14627-0171, USA}\\
      \vspace{0.5cm}

\end{center}

\hspace{5cm}

\begin{abstract}
\noindent
We examine whether the free energy of $\mathcal{N}=4$ super Yang-Mills theory (SYM) in four dimensions corresponds to the partition function of the $AdS_5 \times S^5$ superstring when the corresponding operators are inserted into both theories. We obtain the formal free energy of $\mathcal{N}=4$ $U(N)$ SYM in four dimensions generated by the Feynman graph expansion to all orders of the 't Hooft coupling expansion with arbitrary $N$. This free energy is written as the sum over discretized closed two-dimensional surfaces that are identified with the world-sheets of the string. We compare this free energy with the formal partition function of the discretized $AdS_5 \times S^5$ superstring with the kappa-symmetry fixed in the killing gauge and in the expansion corresponding to the weak 't Hooft coupling expansion in the SYM. We find that some of their properties are identical, although further studies are required to obtain a more precise comparison. Our result suggests a mechanism through which the world-sheet emerges dynamically from $\mathcal{N}=4$ SYM, and this enables us to derive the manner in which the $AdS_5 \times S^5$ superstring is reproduced in the AdS/CFT correspondence.  
\end{abstract}
\vfill
\end{titlepage}
\vfil\eject

\setcounter{footnote}{0}

\section{Introduction}
\setcounter{equation}{0}
The AdS/CFT correspondence proposed by Maldacena states that the superstring theory on the $AdS_5 \times S^5$ background corresponds to $\mathcal{N}=4$ super Yang-Mills theory (SYM) in four dimensions \cite{Maldacena}. In the $g_s \to 0$, $\alpha' \to 0$ limit, it has been established that the supergravity on $AdS_5 \times S^5$ corresponds to $\mathcal{N}=4$ SYM in the large $N$, large 't Hooft coupling ($\lambda=g_{YM}^2N$) limit \cite{{GKP},{Witten},{ST},{AGMOO-DF}}. In the $g_s \to 0$ limit and in the weak $\alpha'$ expansion, which corresponds to the strong $\lambda$ expansion in the SYM ($\lambda=\frac{R^4}{\alpha'^2}$), carrying out comparison is difficult, because we know little about the strong coupling SYM, while we know much about the weak coupling SYM. If we take the BMN limit\footnote{Quantum analysis of the full $AdS_5 \times S^5$ superstring has not yet succeeded, although integrable properties in the superstring are expected to contribute to the analysis \cite{{BenaPolchinskiRoiban},{HatsudaYoshida},{Das},{MannPolchinski}}.} further, the AdS/CFT correspondence reduces to the correspondence between the superstring theory on the pp-wave background and the BMN sector of $\mathcal{N}=4$ SYM in the large $N$ limit \cite{{BerensteinMaldacenaNastase},{MinahanZarembo},{BeisertKristjansenStaudacher},{ArutyunovFrolovRussoTseytlin}}. Although we can compare this limiting string theory with the weak coupling SYM effectively by using the composite coupling expansion ($\lambda/J^2$, where $J$ is the large $R$ charge), we fail to see the correspondence at three loops \cite{{CLMSSW},{SerbanStaudacher}}. Therefore, we need to find a string description dual to the weak coupling SYM \cite{{Tseytlin},{DharMandalWadia},{AldayDavidGavaNarain1},{AldayDavidGavaNarain2}}.

\begin{figure}[htbp]
\begin{center}
\psfrag{i}{i}
\psfrag{j}{j}
\psfrag{1}{(1)}
\psfrag{2}{(2)}
\includegraphics[height=3cm, keepaspectratio, clip]{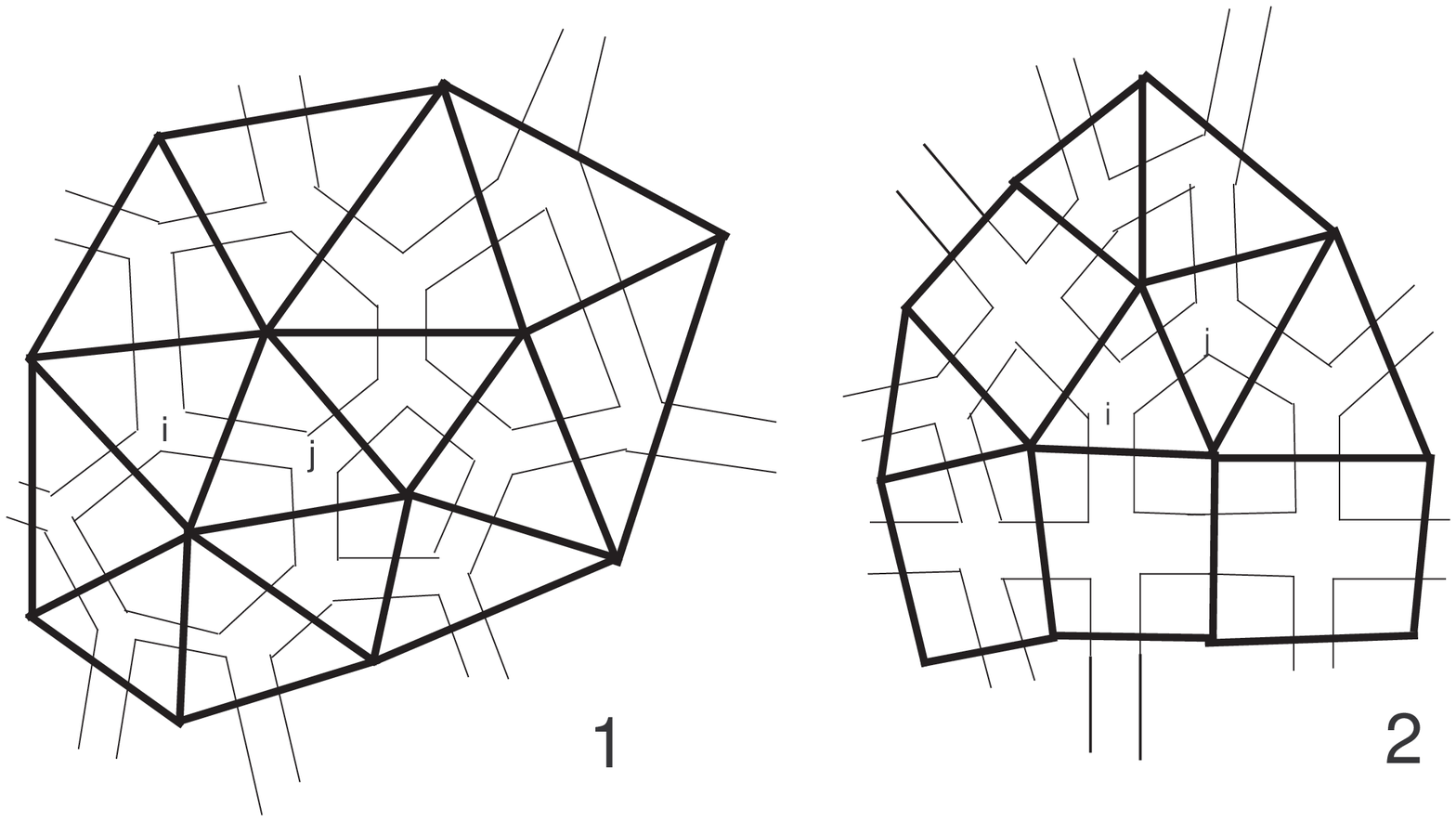}\end{center}
\caption{Random lattices in (1) the \textit{c}=1 matrix model and (2) the SYM. 
}
\label{intro}
\end{figure}

On the other hand, recently the $c \leq 1$ $U(N)$ matrix model, which corresponds to the $c \leq 1$ string theory nonperturbatively,  has been developed extensively \cite{{McGreevyVerlinde},{TakayanagiToumbas},{DKKMMS},{HHIKKMT},{TakayamaTsuchiya}} by incorporating the ideas employed in study of the tachyon condensation \cite{{Sen},{Sato}}, D-branes and the AdS/CFT correspondence. Next, we need to develop the AdS/CFT correspondence by incorporating the ideas used in the $c \leq 1$ matrix model \cite{{BrezinItzyksonParisiZuber},{DistlerKawai},{DasJevicki},{GrossMiljkovic},{GinspargZinn-Justin},{Itoyama},{GinspargMoore}}. One of the most important facts concerning the correspondence between the $c \leq 1$ matrix model and the $c \leq 1$ string theory is that the free energy of the matrix model is identical to the partition function of the string theory. The first piece of evidences supporting this fact is based on the emergence of the world-sheet in the $c \leq 1$ matrix model, which can be summarized as follows \cite{{KazakovMigdal}}. The formal free energy of the matrix model can be written perturbatively as the sum of the connected bubble diagrams. In the double-line notation, each diagram has a dual diagram, as shown in Fig. \ref{intro}, which forms a two-dimensional surface corresponding to the world-sheet of the string. The $N$ dependence of the bubble diagram is $N^{2-2h}$, where $h$ is the number of genuses of the two-dimensional surface. This dependence implies a genus expansion. Consequently, $1/N$ corresponds to the string coupling. Each coefficient of $N^{2-2h}$ corresponds to the formal partition function of the $c \leq 1$ string theory discretized on random lattices with a fixed genus number. This relation shows the correspondence to all orders of not only the $\alpha'$ expansion but also the $g_s$ expansion.

The same kind of relation should also exist in the AdS/CFT correspondence. In fact, the genus expansion was originally found in Yang-Mills theory by 't Hooft \cite{{Hooft}}. We describe it in the following. The formal free energy in Yang-Mills theory can be written perturbatively as the sum of the bubble diagrams, which correspond to world-sheets as in the $c \leq 1$ matrix model (Fig. \ref{intro}). The $N$ and $g_{YM}$ dependence of each bubble diagram are those of $(\frac{1}{g_{YM}^2})^F (g_{YM}^2)^E N^V$, where $F$, $E$ and $V$ are the numbers of faces, edges and vertices in the dual diagram, respectively. If we rewrite $g_{YM}$ as $\lambda = g_{YM}^2 N$ (the 't Hooft coupling), the genus expansion $N^{F-E+V} \lambda^{E-F}= N^{2-2h}\lambda^{E-F}$ is realized. In the AdS/CFT correspondence, $N$ is identified with $\lambda/g_s$, because of the open-closed duality $\lambda = g_{YM}^2 N=g_s N$. By rewriting $N$ in terms of $g_s$, we obtain the dependence of the bubble diagram as $(N/\lambda)^{2-2h}\lambda^V=g_s^{2h-2}\lambda^V$. This dependence indicates the existence of a string theory dual to the weak coupling SYM and a relation in the AdS/CFT correspondence analogous to the correspondence between the free energy of the $c \leq 1$ matrix and the partition function of the $c \leq 1$ string.

The purpose of this paper is to examine whether the free energy of $\mathcal{N}=4$ SYM in four dimensions corresponds to the partition function of the $AdS_5 \times S^5$ superstring to all orders of the string coupling constant when the corresponding operators are inserted into both theories in the same way as in the $c \leq 1$ case\footnote{
This kind of generalization was originally considered in \cite{NastaseSiegel}. We consider this work in section four.}.
 First, we obtain the formal free energy of the SYM generated by the Feynman graph expansion to all orders of the weak coupling $\lambda$ with arbitrary $N$. Next, we attempt to elucidate the string theory dual to the weak coupling SYM. We also obtain a formal partition function of the discretized string theory in the expansion corresponding to the weak coupling expansion of the SYM and compare it with the free energy. We find that the formal partition function and the formal free energy have non-trivial forms and seem to correspond to each other. Moreover, even if we insert certain operators into the formal partition function of the string theory and into the formal free energy of the SYM, this correspondence is still valid. 

Here we comment on the relation between our work and Gopakumar's works on the free $\mathcal{N}$ = 4 SYM, which is  another approach to the AdS/CFT correspondence at the string level \cite{Gopakumar123}. We start with two approaches to the $c \leq 1$  string theory, generalized Kontsevich model (GKM) and the $c \leq 1$ Hermitian matrix model (HMM). Although they are equivalent \cite{{Witten2d},{Kontsevich},{KMMMZ}}, we can easily find the moduli of the string theory in the GKM, while we can easily find the classical action of the string theory in the HMM (see (\ref{c=1 matrix})). Gopakumar treated the free $\mathcal{N}$ = 4 SYM as the GKM and derived the moduli of the $AdS_5 \times S^5$ superstring theory. Three-point and four-point correlation functions have also been obtained in the free $\mathcal{N}$ = 4 SYM and their duals have been proposed \cite{{Gopakumar123},{AKR},{Gopakumar4}}. The thermal case is studied in \cite{Furuuchi}. Because we know little about quantum properties of the string theory, we cannot verify Gopakumar's result at present. Therefore, we treat the $\mathcal{N}$ = 4 SYM in the same manner as the HMM and attempt to derive the classical action of the string theory. \footnote{GKM and HMM for the $AdS_5 \times S^5$ superstring should coincide because of the strong symmetry on both sides in the AdS/CFT correspondence.}

The organization of this paper is as follows. In section 2 we review the equivalence between the formal free energy in the \textit{c}=1 matrix model generated by the Feynman graph expansion and the formal partition function of the \textit{c}=1 string theory discretized in random lattices. In section 3 we obtain the formal free energy of $U(N)$ $\mathcal{N}=4$ SYM in four dimensions generated by the Feynman graph expansion. In section 4 we obtain the formal partition function of the discretized $AdS_5 \times S^5$ superstring in the expansion corresponding to the weak 't Hooft coupling expansion in the SYM and compare it with the free energy obtained in the previous section. In section 5 we summarize and discuss the results. In Appendix A we summarize the Feynman rules for $\mathcal{N}=4$ $U(N)$ SYM in the double-line notation. In Appendix B we obtain the formal free energy of ten-dimensional $U(N)$ $\mathcal{N}=1$ SYM in the same manner as in the four-dimensional $\mathcal{N}=4$ SYM case.

\section{\textit{c}=1 Matrix Model and Discretized \textit{c}=1 String Theory}
\setcounter{equation}{0}
In this section, we briefly review the manner in which the world-sheets emerge in the \textit{c}=1 matrix model. In particular, we show that the perturbative formal free energy of the \textit{c}=1 matrix model is equivalent to the formal partition function of the \textit{c}=1 string theory discretized on random lattices.

\begin{figure}[htbp]
\begin{center}
\psfrag{Ng}{Ng}
\includegraphics[height=3cm, keepaspectratio, clip]{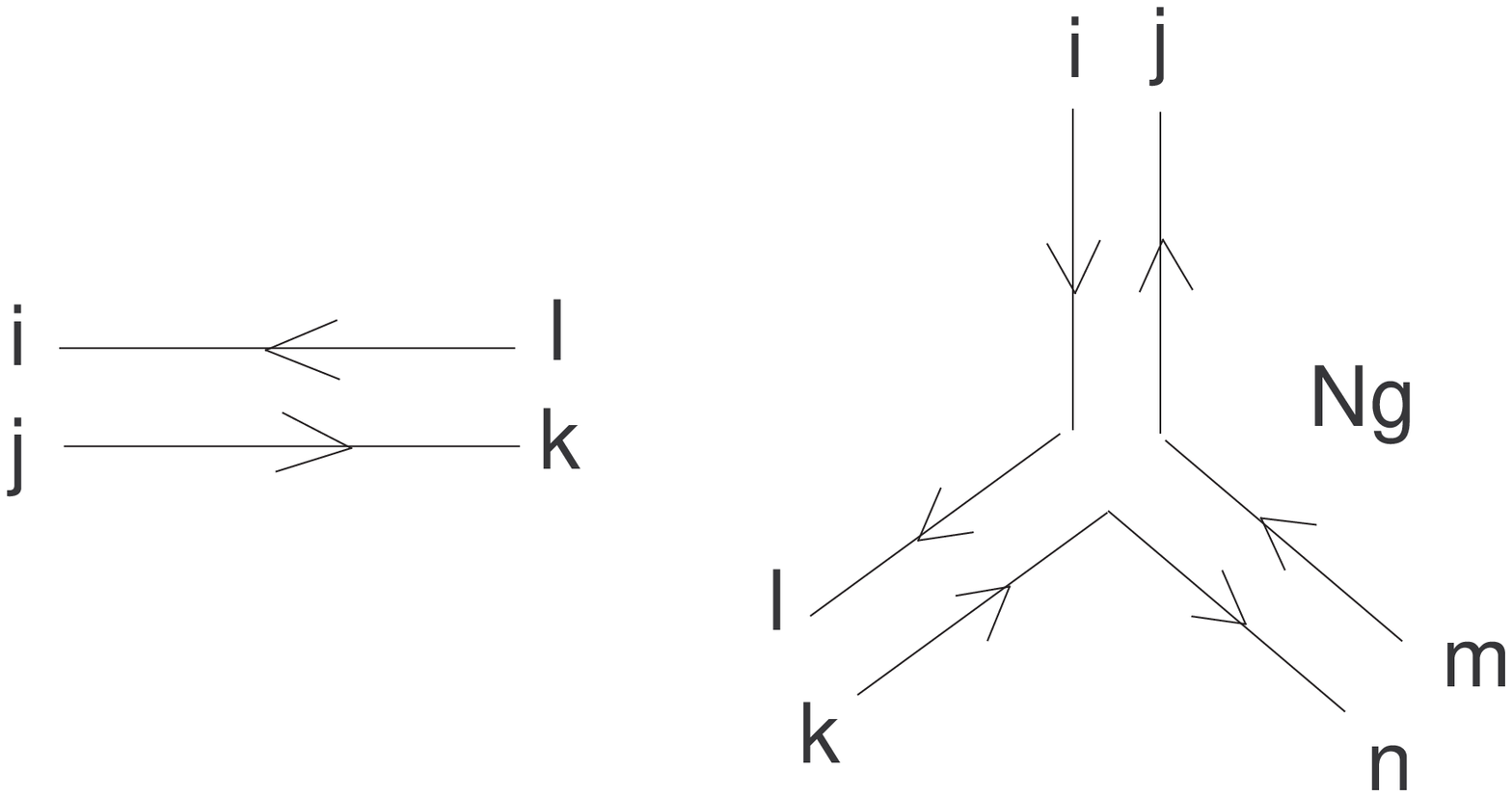}\end{center}
\caption{Feynman rules for the \textit{c}=1 matrix model.}
\label{c=1rule}
\end{figure}

We begin with the action of the \textit{c}=1 matrix model,
\beq
Z_{\mbox{matrix}}
=
N\int^{\infty}_{-\infty}dt 
\tr\left(\frac{1}{2}(\partial_t\Phi)^2+\frac{1}{2\alpha'^2}\Phi^2-\frac{1}{3}g\Phi^3 \right),
\eeq
where $\Phi$ is a $N \times N$ Hermitian matrix.
The propagator in the double-line notation is given by
\beq
\langle \Phi_{ij}(t)\Phi_{kl}(t') \rangle =\frac{1}{N}\Delta(t-t')\delta_{il}\delta_{jk},
\eeq
where
\beq
\Delta(t-t')
=
\int^{\infty} _{-\infty} dk \frac{1}{k^2+\frac{1}{\alpha'^2}}e^{ik(t-t')}
=
e^{-\frac{1}{\alpha'}|t-t'|},
\eeq
whereas the interaction vertex is given by 
\beq
Ng\delta_{il}\delta_{kn}\delta_{mj}.
\eeq
The Feynman diagrams are shown in Fig. \ref{c=1rule}.

The formal free energy $F_{\mbox{matrix}}$ can be written perturbatively as the sum of the connected bubble diagrams. The $N$ dependence of each diagram is given by 
$N^{\mid\mbox{interaction vertices}\mid}\times 1/N^{\mid\mbox{propagators}\mid}\times N^{\mid\mbox{inner loops}\mid}$, where $\mid\mbox{propagators}\mid$ for example represents the number of propagators. Each diagram forms a two-dimensional closed surface, which is represented by double lines in Fig. \ref{intro}. The diagram has a dual diagram, which is represented by bold lines in Fig. \ref{intro}; the interaction vertices, propagators and inner loops correspond to the faces, edges and vertices in the dual diagram, respectively. Then the dual diagram forms a random lattice on the two-dimensional closed surface that is identified with the world-sheet of the dual string below. If we represent the number of the faces, edges and vertices in the random lattice as $F$, $E$ and $V$, respectively, the $N$ dependence of the Feynman diagram is expressed as $N^{F-E+V}=N^{2-2h}$, where $h$ is the number of the genuses of the random lattice. Therefore, $F_{\mbox{matrix}}$ is given by the following genus expansion,
\beq
F_{\mbox{matrix}}
=
\sum_{\mbox{random lattices}} N^{2-2h} g^F \prod_{k=1}^F
\int d t_k \exp \left(\sum_{\langle i,j \rangle (i<j)}-\frac{1}{\alpha'}|t_i-t_j| \right),
\label{c=1 matrix}
\eeq
where $i$, $j$ and $k$ represent the points of the interaction vertices in the Feynman diagrams, and $\langle i,j\rangle  $ represent the nearest neighbors among the interaction vertices, as can be seen in Fig. \ref{intro}. The summation $\sum_{\mbox{random lattices}}$ is that over arbitrary random lattices that form two-dimensional closed surfaces. We ignore the symmetric factors, which go to 1 in the continuum limit. 

Next, let us consider the \textit{c}=1 string theory. By discretizing the partition function
\beq
Z_{\mbox{c=1 string}}
=
\sum_h \int Dg \int Dt 
\exp\left[
-\gamma \frac{1}{4\pi} \int d^2 \sigma \sqrt{g}R 
-\beta \int d^2 \sigma \sqrt{g}
-\frac{1}{2 \pi \alpha'} \int d^2 \sigma \sqrt{g}\partial_{\alpha}t\partial^{\alpha}t
\right] \label{c=1string}
\eeq
on random lattices by taking
\beqa
\sum_h \int Dg &\to& \sum_{\mbox{random lattices}} \n
\frac{1}{2 \pi } \int d^2 \sigma \sqrt{g} \partial_{\alpha}t\partial^{\alpha}t &\to& \sum_{\langle i,j\rangle  (i<j)} (t_i-t_j)^2
\eeqa 
and by making the identification $e^{\gamma} \equiv 1/N$ and $e^{-\beta A} \equiv g$, where $A$ is the area of one face $(AF=\int d^2 \sigma \sqrt{g})$, we obtain
\beq
Z_{\mbox{c=1 string}}
=
\sum_{\mbox{random lattices}} N^{2-2h} g^F \prod_{k=1}^F
\int d t_k \exp \left(\sum_{\langle i,j\rangle  (i<j)}-\frac{1}{\alpha'}(t_i-t_j)^2 \right).
\label{c=1disc}
\eeq
This theory and the theory whose formal partition function takes the form (\ref{c=1 matrix}) belong to the same universality class in the continuum limit. Therefore, the perturbative formal free energy of the \textit{c}=1 matrix model is equivalent to the formal partition function of the \textit{c}=1 string theory discretized on random lattices.

This is the first piece of evidences supporting the correspondence between the \textit{c}=1 matrix model and the \textit{c}=1 string theory to all orders of the string coupling and $\alpha'$ coupling.

\vspace{1cm}

\section{$\mathcal{N}=4$ Super Yang-Mills}
\setcounter{equation}{0}

In this section, we obtain the formal free energy of $\mathcal{N}=4$ $U(N)$ super Yang-Mills theory (SYM) in four dimensions analogous to (\ref{c=1 matrix}). 

We begin with the action of $\mathcal{N}=4$ $U(N)$ SYM in the Feynman gauge,
\beqa
S_{SYM}&=&\frac{1}{2 g_{YM}^2}\int d^4x 
\tr 
\left( -\frac{1}{2}F^{\mu\nu}F_{\mu\nu} -(\partial_{\mu}A^\mu)^2
-D^{\mu}X^I D_{\mu}X_I +\frac{1}{2} [X^I, X^J] [X_I, X_J]
+ 2\bar{c}\partial^{\mu}D_{\mu}c \right. \n
&& \qquad \qquad \qquad \quad +  \psi^T i\gamma^{\mu} D_{\mu} \psi
+ \psi^T \gamma^I [X_I, \psi] \Biggr), 
\eeqa   
where $\mu$ and $I$ run from 0 to 3 and from 4 to 9, respectively, and 
$D_{\mu}=\partial_{\mu}-i[A_{\mu},\;\;]$. All fields are represented by Hermitian $N \times N$ matrices. We use the Majorana-Weyl representation in ten dimensions, where $\gamma^M$ and $\hat{\gamma}^M$ $(M=0, \cdots, 9)$ are 16 $\times$ 16 matrices that satisfy the following properties:
\beqa
&&\gamma^M=(\gamma^{\mu}, \gamma^I)=(-1, \gamma^i), \quad 
\hat{\gamma}^M=(\hat{\gamma}^{\mu}, \hat{\gamma}^I)=(1, \gamma^i) \quad
(i=1, \cdots, 9) \n
&& \{\gamma^i, \gamma^j\}=2 \delta^{ij}, \quad \gamma^{iT}=\gamma^i, 
\label{gamma}
\eeqa
from which $\gamma^M\hat{\gamma}^N+\gamma^N\hat{\gamma}^M=2\eta^{MN}$ and
$\hat{\gamma}^M\gamma^N+\hat{\gamma}^N\gamma^M=2\eta^{MN}$ are satisfied.
$\psi$ is a 16 component Weyl spinor that satisfies the Majorana condition
$\psi^*=\psi$. 
We summarize the Feynman rules in the double-line notation in Appendix A.

\subsection{Formal Free Energy in Bosonic Part of $\mathcal{N}=4$ SYM}

In this subsection we obtain the formal free energy of the bosonic part of $\mathcal{N}=4$ SYM to all orders of $\lambda$ in the Feynman graph expansion with arbitrary $N$. In the following we consequentially sum up the interactions in order to explain our result. 

\begin{figure}[htbp]
\begin{center}
\psfrag{i}{i}
\psfrag{j}{j}
\psfrag{M}{$M_{ij}$}
\psfrag{N}{$M_{ji}$}
\psfrag{a}{$A^{\mu}$}
\psfrag{b}{$X^I$}
\psfrag{c}{$c$}
\psfrag{d}{$\bar{c}$}
\psfrag{e}{$\psi$}
\psfrag{f}{$\hat{\psi}$}
\psfrag{0}{$0, \cdots, 3$}
\psfrag{4}{$4, \cdots, 9$}
\psfrag{10}{10}
\psfrag{11}{11}
\psfrag{12}{12}
\psfrag{13}{13}
\psfrag{14}{14}
\psfrag{15}{15}
\includegraphics[height=3.5cm, keepaspectratio, clip]{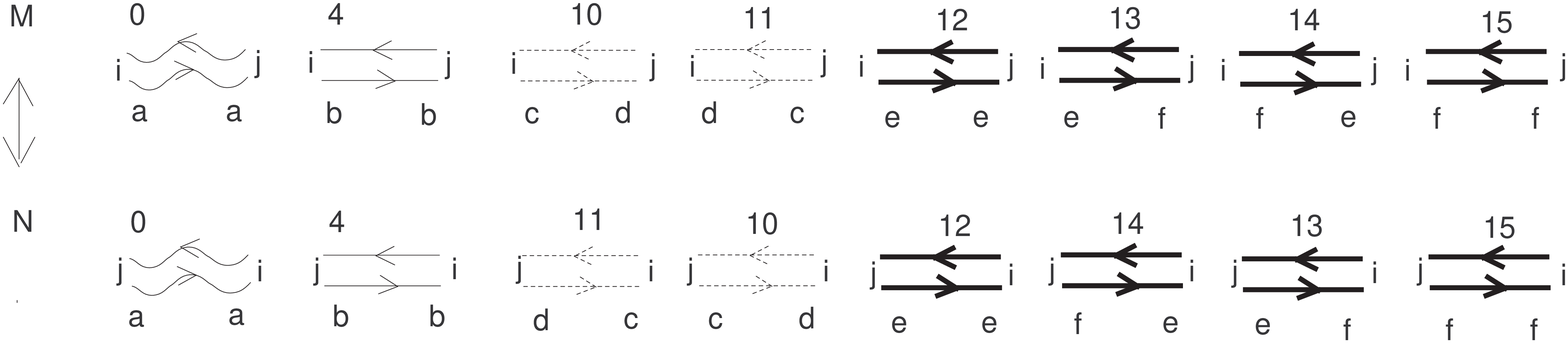}\end{center}
\caption{The quantum numbers on links.}
\label{quantum}
\end{figure}

\begin{figure}[htbp]
\begin{center}
\psfrag{i}{i}
\psfrag{j1}{$j_1$}
\psfrag{j2}{$j_2$}
\psfrag{j3}{$j_3$}
\psfrag{j4}{$j_4$}
\psfrag{M1}{$M_{ij_1}$}
\psfrag{M2}{$M_{ij_2}$}
\psfrag{M3}{$M_{ij_3}$}
\psfrag{M4}{$M_{ij_4}$}
\psfrag{k1}{$k_{ij_1}$}
\psfrag{k2}{$k_{ij_2}$}
\psfrag{k3}{$k_{ij_3}$}
\psfrag{k4}{$k_{ij_4}$}
\includegraphics[height=5cm, keepaspectratio, clip]{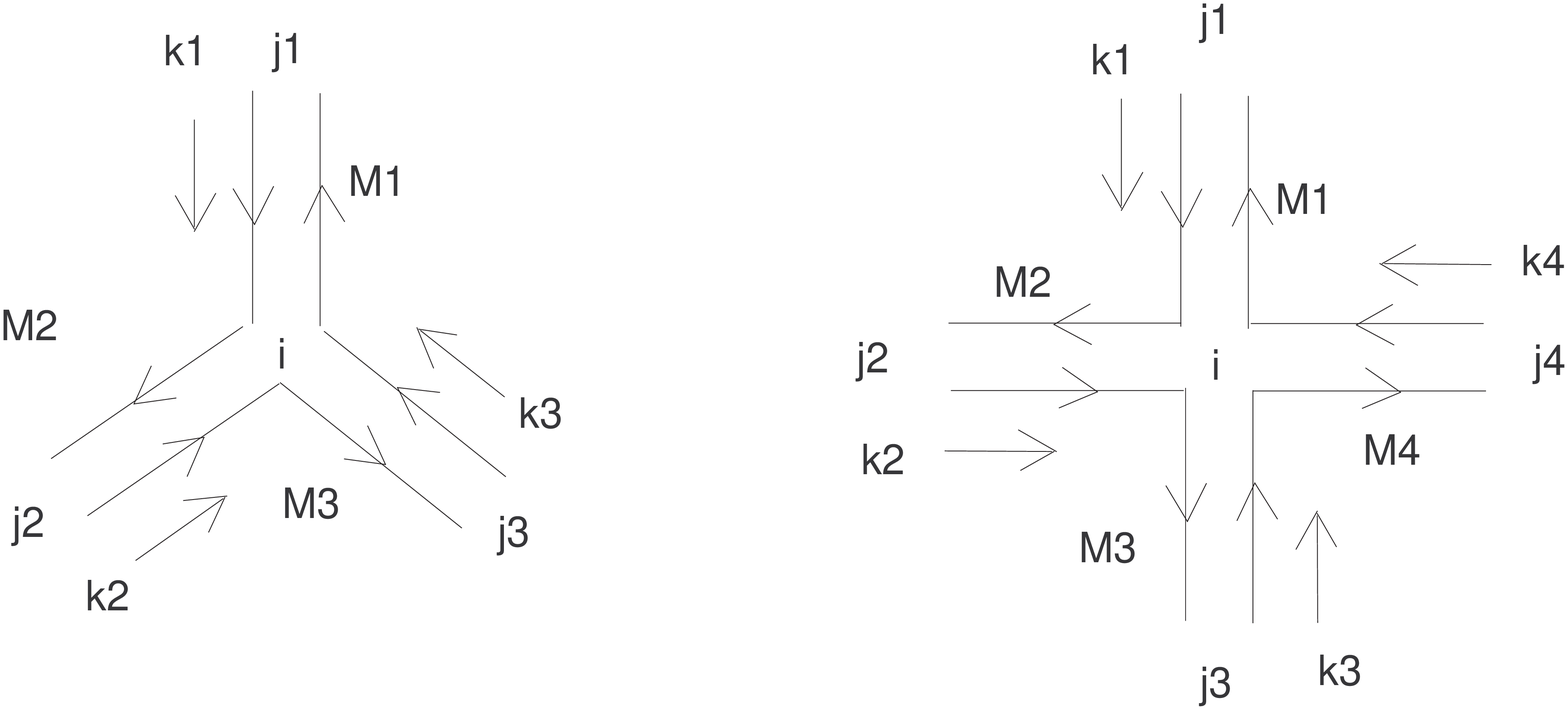}\end{center}
\caption{Notation for the three- and four-point interactions.}
\label{3pt}
\end{figure}

The difficulty encountered in attempting to derive the free energy is overcome by assigning quantum numbers $M_{ij}$ ($=0, \cdots, 15$) that represent the propagators on the links between the vertices $i$ and $j$. $M_{ij}$ and the relations between $M_{ij}$ and $M_{ji}$ are defined in Fig. \ref{quantum}. First, we sum up the three-point interactions among the gauge fields $A^{\mu}$ and the scalar fields $X^I$ appearing in Fig. \ref{3ptfeynman} in Appendix A. The result is \footnote{We redefine the -i$g_{YM}^2$ to $g_{YM}^2$.}
\beqa
F_{\mbox{3pt.}(A,\; X)}
&=&
\sum_{\mbox{random lattices}}\sum_{M_{ij}=0, \cdots, 9}N^{2-2h}\lambda^{E-F}
\prod_{\langle i,j\rangle  (i<j)} 
\left(
\int d^{10} k_{ij} \delta^6(k^I_{ij}) \Delta(k^{\mu}_{ij})
\right) \n
&& \qquad \prod_i 
\Biggl[ 
\delta^4(\sum_{q=1}^3 k_{ij_q}) 
\Bigl(\eta_{M_{ij_1}M_{ij_2}}(k_{ij_1}^{M_{ij_3}}-k_{ij_2}^{M_{ij_3}})\Bigr)
+\Bigl(\mbox{cyclic in (1,2,3)}\Bigr)
\Biggr], \label{3ptAX}
\eeqa
where $k^{\mu}_{ij}$ is defined to flow from $j$ to $i$ and to satisfy $k^{\mu}_{ji}= -k^{\mu}_{ij}$. Throughout this paper, we ignore symmetric factors, which go to 1 in the continuum limit. In the second line of (\ref{3ptAX}), $i$ represents vertices and $j_q$ $(q=1,2,3)$ represents the vertices in the $i$ counterclockwise direction (see Fig. \ref{3pt}). In this formula, we define $k^I_{ij}$ $(I=4, \cdots, 9)$ and insert $\delta^6(k^I_{ij})$ in order to sum up the $AAA$ interactions and the $AXX$ interactions. For example, when $M_{ij_3}=4, \cdots, 9$, the first term in the second line does not contribute, whereas when  $M_{ij_3}=0, \cdots, 3$, it does contribute.

In the same way, we obtain the following form in the sector of gauge fields $A^{\mu}$ and ghosts $c$ and $\bar{c}$:
\beqa
F_{\mbox{3pt.}(A,\; c, \;\bar{c})}
&=&
\sum_{\mbox{random lattices}}\sum_{M_{ij}=0, \cdots, 3, 10,11}N^{2-2h}\lambda^{E-F}(-1)^{\mid\mbox{fermion loops}\mid} \n
&&
\prod_{\langle i,j\rangle  (i<j)} 
\left(
\int d^{6} k_{ij} \delta^2(k^a_{ij}) \Delta(k^{\mu}_{ij})
\right) \n
&& \prod_i 
\Biggl[\delta^4(\sum_{q=1}^3k_{ij_q}) 
\eta_{M_{ij_1}M_{j_2i}}\Bigl((1-\delta_{M_{ij_1}, 10})k_{ij_1}^{M_{ij_3}}-(1-\delta_{M_{ij_2}, 10})k_{ij_2}^{M_{ij_3}}\Bigr) +\left(\mbox{cyclic in (1,2,3)}\right)
\Biggr], \nonumber
\eeqa
where we have introduced $k^a_{ij}$ $(a=10, 11)$ and inserted $\delta^2(k^a_{ij})$ in order to sum up the $AAA$ interactions and the $Ac\bar{c}$ ($Ac\bar{c}$) interactions.

Finally, in the full bosonic sector of $\mathcal{N}=4$ SYM, we need to sum over arbitrary random lattices that consist of three-point vertices $i(3pt.)$ and four-point vertices $i(4pt.)$, as shown in  Figs. \ref{intro} and \ref{3pt}. The free energy is given by 
\beqa
\!\!\!\!F_{\mbox{bosonic}}
\!\!\!\!&=& \!\!\!\!\!\!\!\!
\sum_{\mbox{random lattices}}\sum_{M_{ij}=0, \cdots, 11}N^{2-2h}\lambda^{E-F}(-1)^{\mid\mbox{fermion loops}\mid} \n
&& \!\!\!\!\!\!\!\!\!\!
\prod_{\langle i,j\rangle  (i<j)} 
 \left(
\int d^{12} k_{ij} \delta^6(k^I_{ij})\delta^2(k^a_{ij})
\Delta(k^{\mu}_{ij})
\right) \n 
\!\!\!\!\!\!\!\!\!\!\!\!&&\!\!\!\!\!\!\!\!\!\!\!\!\!\!\!\!\!\!\prod_{i(4pt.)}
 \left[
\delta^4(\sum_{p=1}^4k_{ij_p})
\prod_{t=10}^{11}\prod_{e=1}^4(1-\delta_{M_{ij_e},t}) 
(2\eta_{M_{ij_1}M_{ij_3}}\eta_{M_{ij_2}M_{ij_4}}-\eta_{M_{ij_1}M_{ij_2}}\eta_{M_{ij_3}M_{ij_4}}-\eta_{M_{ij_1}M_{ij_4}}\eta_{M_{ij_2}M_{ij_3}})\right]
\n
\!\!\!\!\!\!\!\!&& \!\!\!\!\!\!\!\!\!\!\!\!\!\!\!\!\!\! \prod_{i(3pt.)}
\Biggl[  
\delta^4(\sum_{q=1}^3k_{ij_q}) \left(\eta_{M_{ij_1}M_{j_{2}i}}\Bigl((1-\delta_{M_{ij_1}, 10})k_{ij_1}^{M_{ij_3}}-(1-\delta_{M_{ij_2}, 10})k_{ij_2}^{M_{ij_3}}\Bigr) \!+\!\left(\mbox{cyclic in (1,2,3)}\right)\right) 
\Biggr], \nonumber
\eeqa
where we have used the projection $\prod_{t=10}^{11}\prod_{e=1}^4(1-\delta_{M_{ij_e},t})$, because there is no ghost contribution to the four-point vertices. 

\subsection{Formal Free Energy of $\mathcal{N}=4$ SYM}
\begin{figure}[htbp]
\begin{center}
\psfrag{a1}{$A^{\mu}$}
\psfrag{a2}{$A^{\nu}$}
\psfrag{f}{$\psi$}
\psfrag{f'}{$\hat{\psi}$}
\includegraphics[height=2cm, keepaspectratio, clip]{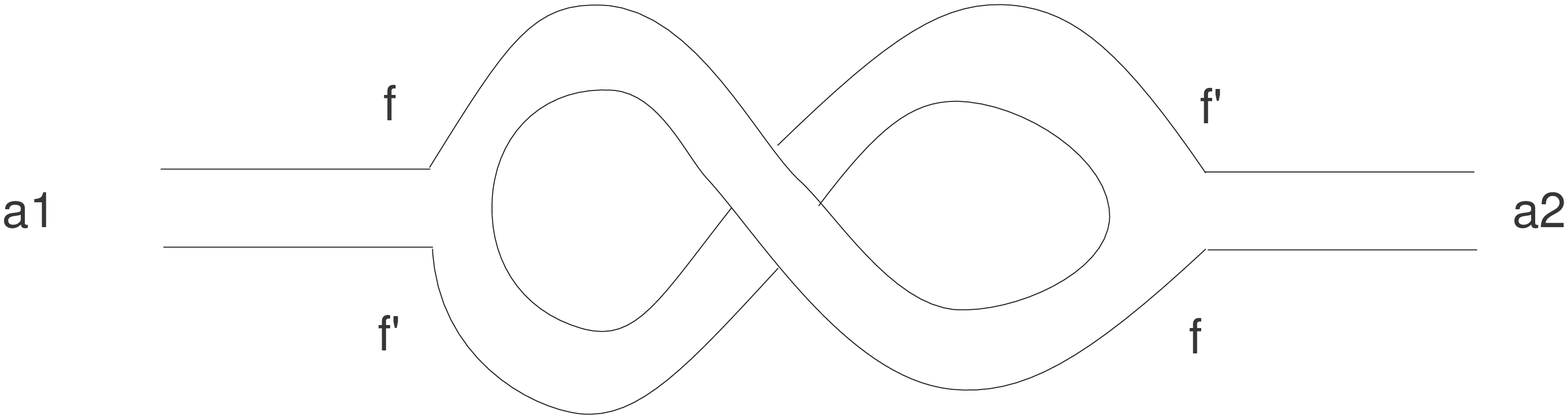}\end{center}
\caption{An example.}
\label{nonplanar}
\end{figure}

In this subsection, we incorporate the fermion interactions and obtain the formal free energy of $\mathcal{N}=4$ SYM. In general, we need to take account of the minus signs that result from interchanges of Majorana fermions, in addition to counting the number of fermion loops, because there are two ways of performing Wick contractions:
$\wick{11}{<1\psi \int d^4x tr( >1\psi^T \gamma^I [X_I, <2\psi]) 
>2\psi}$ and 
$\wick{12}{<1\psi \int d^4x tr( <2\psi^T \gamma^I [X_I, >1\psi]) 
>2\psi}$. In the ordinary notation of the Feynman diagrams, we do not need to consider the minus signs, except for that for the fermion loops, because of the following property:
\beq
\wick{11}{<1\psi^a 
\int d^4x ( >1\psi^{pT} \gamma^I X_I^q <2\psi^r)
tr(t^p [t^q, t^r]) 
>2\psi^b}
=
\wick{12}{<1\psi^a
 \int d^4x ( <2\psi^{pT} \gamma^I X_I^q >1\psi^r) 
tr(t^p [t^q, t^r])
>2\psi^b}.
\eeq
However, in the double-line notation, there is no such property:
\beq
\wick{11}{<1\psi_{ij} \int d^4x 2tr( >1\psi^T \gamma^I X_I <2\psi) 
>2\psi_{kl}}
\neq
\wick{12}{<1\psi_{ij} \int d^4x 2tr( <2\psi^T \gamma^I X_I >1\psi) 
>2\psi_{kl}}.
\eeq
 Therefore, we introduce $\hat{\psi}$ in the action to count the minus signs corresponding to fermion interchanges as follows:
\beqa
&&\frac{1}{2 g_{YM}}\int d^4x 
\tr 
(  \psi^T i\gamma^{\mu} \partial_{\mu} \psi
+ 2\psi^T \gamma^{\mu} A_{\mu} \psi  
+ 2\psi^T \gamma^I X_I \psi)\n
&\to& 
\frac{1}{2 g_{YM}}\int d^4x 
\tr 
(  \hat{\psi}^T i\gamma^{\mu} \partial_{\mu} \psi
+ 2\hat{\psi}^T \gamma^{\mu} A_{\mu} \psi  
+ 2\hat{\psi}^T \gamma^I X_I \psi). \label{modification}
\eeqa
The Feynman diagrams are given in (\ref{SYMpropagator}) and Fig. \ref{3ptfeynman}. We can count the minus signs in the vacuum diagrams by flipping the fermions to the standard position as $\hat{\psi} \psi \hat{\psi} \psi \hat{\psi} \psi \cdots $. For example, in the diagram in Fig. \ref{nonplanar} $\mid$fermion interchanges$\mid$=1 and  $\mid$fermion loops$\mid$=1. 

The formal free energy of $\mathcal{N}=4$ $U(N)$ SYM is perturbatively given by
\beqa
\!\!\!\!F_{\mbox{SYM}}
\!\!\!\!&=& \!\!\!\!\!\!\!\!
\sum_{\mbox{random lattices}}\sum_{M_{ij}=0, \cdots, 15}N^{2-2h}\lambda^{E-F}(-1)^{\mid\mbox{fermion loops}\mid+\mid\mbox{fermion interchanges}\mid} \n
&& \!\!\!\!\!\!\!\!\!\!
\prod_{\langle i,j\rangle  (i<j)} 
\left[
\int d^{12} k_{ij} \delta^6(k^I_{ij})\delta^2(k^a_{ij})
\left(
\sum_{s=12}^{15}(1-\delta_{M_{ij},s})\Delta(k^{\mu}_{ij})
+\sum_{s=12}^{15}\delta_{M_{ij},s}\Delta_s^{ij}(k^{\mu}_{ij})
\right)
\right] \n 
\!\!\!\!&&\!\!\!\!\!\!\!\!\!\!
\prod_{i(4pt.)}
 \left[
\delta^4(\sum_{p=1}^4k_{ij_p})
\prod_{t=10}^{15}\prod_{e=1}^4(1-\delta_{M_{ij_e},t}) 
(2\eta_{M_{ij_1}M_{ij_3}}\eta_{M_{ij_2}M_{ij_4}}-\eta_{M_{ij_1}M_{ij_2}}\eta_{M_{ij_3}M_{ij_4}}-\eta_{M_{ij_1}M_{ij_4}}\eta_{M_{ij_2}M_{ij_3}})
\right]
\n
&& \!\!\!\!\!\!\!\!\!\! \prod_{i(3pt.)}
\delta^4(\sum_{q=1}^3k_{ij_q})
\Biggl[
\prod_{u=12}^{15}\prod_{f=1}^3(1-\delta_{M_{ij_f},u})  
\Biggl( \eta_{M_{ij_1}M_{j_2i}}\Bigl((1-\delta_{M_{ij_1}, 10})k_{ij_1}^{M_{ij_3}}-(1-\delta_{M_{ij_2}, 10})k_{ij_2}^{M_{ij_3}}\Bigr)\n&& \qquad\qquad\qquad\qquad\qquad\qquad\qquad\qquad\qquad\qquad\qquad +\left(\mbox{cyclic in (1,2,3)}\right) \Biggr) \n
&& \qquad\qquad\qquad\quad
+\Bigl(
(-1)^{1+\Theta(i-j_2)+\Theta(i-j_3)}
(\delta_{M_{ij_2},12}+\delta_{M_{ij_2},13})
(\delta_{M_{ij_3},14}+\delta_{M_{ij_3},15})
\gamma_i^{M_{ij_1}} \n
&& \qquad\qquad\qquad\qquad\qquad\qquad\qquad\qquad\qquad\qquad\qquad
+\left(\mbox{cyclic in (1,2,3)}\right)
\Biggr)
\Biggr], \label{SYMresult}
\eeqa
where we assign the bosonic propagators (\ref{bosonicpropagator}) to the links with $M_{ij}=0, \cdots, 11$  and the fermionic propagators (\ref{fermionicpropagator}) to the links with $M_{ij}=12, \cdots, 15$.
We formally define $\gamma^M=0$ in the cases $M=10, \cdots, 15$. The subscripts $i$ and $j$ of $\Delta_{ij}$ and the subscript $i$ of $\gamma^M_i$ indicate the order of the multiplication of the gamma matrices. For example, $\Delta_{i_1 i_2}(k_{i_1 i_2}) \gamma^{M_2}_{i_2} \Delta_{i_2 i_3}(k_{i_2 i_3}) \gamma^{M_3}_{i_3} \Delta_{i_3 i_1}(k_{i_3 i_1}) \gamma^{M_1}_{i_1} $ represents $Tr(\Delta(k_{i_1 i_2}) \gamma^{M_2} \Delta(k_{i_2 i_3}) \gamma^{M_3} \Delta(k_{i_3 i_1}) \gamma^{M_1} )$, where $\Tr$ represents the trace over the spinor indices. The directions of the fermionic propagators $\Delta_s(k_{ij}^{\mu})$ $(i<j)$ produce the factors $(-1)^{1+\Theta(i-j_2)+\Theta(i-j_3)}$
in the sixth line of (\ref{SYMresult}) and their permutations, where $\theta(i,j)=1$ $(i>j)$ or $=0$ $(i<j)$. 
We note that $N^{2-2h}\lambda^{E-F}$ in the above formula can be rewritten as $g_s^{2h-2}\lambda^V$ in terms of the closed string, as discussed in the introduction.\footnote{
In the \textit{c}=1 case, the dependence $g^F$ of the free energy of the matrix model implies the existence of the cosmological constant in the dual string. By contrast, the fact that there is no such dependence of the free energy for the SYM implies that there is no cosmological constant in the $AdS_5 \times S^5$ superstring. This is consistent with the Weyl invariance in the $AdS_5 \times S^5$ superstring.
}

In the next section, we examine whether this free energy of $\mathcal{N}=4$ SYM is equivalent to the formal partition function of the $AdS_5 \times S^5$ superstring discretized on random lattices. A problem that we immediately recognize is that there are only four continuous degrees of freedom in the formal free energy, which are integrals over $k^{\mu}$, whereas there are ten scalar degrees of freedom and the fermionic degrees of freedom in the formal partition function of the string. In the next section, we find a natural solution to this problem by considering the string theory in the limit corresponding to the weak 't Hooft coupling limit in the SYM.


\vspace{1cm}

\section{Discretized $AdS_5 \times S^5$ Superstring}
\setcounter{equation}{0}
In the previous section, we obtained the formal free energy of the SYM by carrying out the weak 't Hooft coupling expansion. In this section, we consider, for comparison, the $AdS_5 \times S^5$ superstring dual to the weak 't Hooft coupling SYM. The formal partition function of the superstring is then obtained on the discretized world-sheet with the expansion corresponding to the weak 't Hooft coupling expansion in the SYM, and it is compared with the previous result.

In order to understand a proper method for taking a limit in the $AdS_5 \times S^5$ superstring, which corresponds to the weak coupling limit in the SYM, we review the original observation of $N$ coincident D3-branes. The conjecture regarding the AdS/CFT correspondence is based on the observation of D3-branes from the open sting and closed string points of view. We can summarize this observation as follows. From the open string point of view, if we observe the vicinity of the D3-branes, this system is described by $\mathcal{N}=4$ $U(N)$ SYM \cite{DKPS}. In contrast, from the closed string point of view, we begin with the $N$ coincident D3-brane background,
\beq
ds^2=f^{-\frac{1}{2}}(dx^{\mu})^2 + f^{\frac{1}{2}}(dr^2 + r^2 (d \Omega_5)^2),
\eeq 
where
\beq
f=1 + \frac{g_s N \alpha'^2}{r^4}.
\eeq
The closed string coupling constant $g_s$ in this formula can be rewritten in terms of the gauge theory by using the relation in the open-closed duality, $g_s N=g_{YM}^2N=\lambda$. In order to observe the vicinity of the D3-branes, we need to rescale the radial coordinate from $r$ to $U$ as
\beq
U:=\frac{r}{\alpha'}
\eeq
and to take the $\alpha' \to 0$ limit (the near horizon limit), with the 't Hooft coupling $\lambda$ fixed. As a result, we obtain the $AdS_5 \times S^5$ background,
\beq
ds^2=\alpha' \left( \frac{U^2}{\sqrt{\lambda}} (dx^{\mu})^2 
+  \frac{\sqrt{\lambda}}{U^2} dU^2 + \sqrt{\lambda} (d \Omega_5)^2 \right).
\eeq
The AdS radius and the $S^5$ radius have the same value $R=\lambda^{\frac{1}{4}}\sqrt{\alpha'}$.
For convenience, we change the coordinate from $U$ to $u$ as $u:= \frac{1}{U}$ and obtain
\beq
ds^2= \frac{\alpha'}{\sqrt{\lambda}} \left(
\frac{((dx^{\mu})^2 + \lambda du^2 )}{u^2}
+ \lambda (d \Omega_5)^2 
\right). \label{background}
\eeq
This observation suggests the correspondence between $\mathcal{N}=4$ $U(N)$ SYM and the superstring on the $AdS_5 \times S^5$ background, (\ref{background}). 

In this full correspondence, there exist degrees of freedom of a coordinate transformation in the target space of the $AdS_5 \times S^5$ superstring when the coupling constant $\lambda$ is finite, because the coordinate transformation is absorbed by the field redefinition of the scalars that represent the coordinates. However, if a limit of $\lambda$ is taken, the theory obviously depends on which coordinates are used \cite{Tseytlin}. Therefore, when we consider the weak or strong coupling limit, we need to fix the coordinates as follows. 

The system dual to the strong coupling $\mathcal{N}=4$ SYM can be described by the supergravity on the corresponding background, because the curvature of the background becomes small as $\lambda = \frac{R^4}{\alpha'^2} \to \infty$. In the strong 't Hooft coupling limit of the background metric (\ref{background}), it shrinks to a six-dimensional metric, and the supergravity description becomes invalid. For this reason, we need to further rescale the coordinate from $u$ to $\tilde{u}$ as $\tilde{u}:=\sqrt{\lambda}u$, and take the limit $\lambda \to \infty$ with the coordinate $\tilde{u}$ fixed. We then obtain 
\beq
ds^2=\alpha' \sqrt{\lambda} \left(
\frac{(dx^{\mu})^2 + d\tilde{u}^2 }{\tilde{u}^2} + (d \Omega_5)^2 
\right). \label{strong}
\eeq
This is the procedure by which the coordinates are fixed on the gravity side in the established duality between the strong coupling $\mathcal{N}=4$ SYM and the type IIB supergravity on the $AdS_5 \times S^5$ background.

Next, in order to obtain the dual system to the weak coupling $\mathcal{N}=4$ SYM, we need to take the opposite limit, i.e. $\lambda (= \frac{R^4}{\alpha'^2})\to 0$ of (\ref{background}). In this case, because the curvature of the background becomes very large, supergravity can no longer describe this system. Therefore, we employ the string description for this system,
\beq
S_{\mbox{boson}}=\frac{1}{\sqrt{\lambda}}
\int d^2 \sigma \frac{1}{2}\sqrt{-g}
\left(
\frac{((\partial_{\alpha}x^{\mu})^2 + \lambda 
(\partial_{\alpha} u)^2 )}{u^2}
+ \lambda(\partial_{\alpha} \Omega_5)^2 
\right). \label{fullboson}
\eeq
By discretizing the world-sheet, we obtain 
\beq
\bar{S}_{\mbox{boson}}=\frac{1}{2\sqrt{\lambda}}
\sum_{\langle i, j \rangle}
\left(
\frac{((x^{\mu}_i-x^{\mu}_j)^2 + \lambda 
(u_i-u_j)^2 )}{u_iu_j}
+ \lambda(\Omega_{5i}-\Omega_{5j})^2 
\right), \label{fullboson}
\eeq
which is defined on a discretized world-sheet that is constructed from triangles and quadrangles. We fix the original coordinates (\ref{background}), because we must have only four-dimensional coordinates in the weak coupling limit in order to reproduce the result from the previous section.
In fact, in the $\lambda \to 0$ limit, (\ref{fullboson}) reduces to the leading-order term on each discretized world-sheet, 
\beq
\bar{S}_{\mbox{boson leading}}=\frac{1}{2\sqrt{\lambda}}
\sum_{\langle i, j \rangle}
\left(
\frac{(x^{\mu}_i-x^{\mu}_j)^2 }{u_iu_j} 
\right).\label{leading} 
\eeq
We treat the other terms in (\ref{fullboson}) as higher-order corrections to the action (\ref{leading}) on each discretized world-sheet in the limit of the small coupling $\lambda$. This weak coupling expansion in the $AdS_5 \times S^5$ superstring was originally considered in \cite{NastaseSiegel}.

To this point, we have discussed the bosonic part of the closed string. In the following, we consider the full description of the $AdS_5 \times S^5$ superstring. Because it is difficult to discretize the superstring action with manifest kappa symmetry \cite{{Siegel},{Ambjorn}}, we start with the $AdS_5 \times S^5$ superstring with the kappa symmetry fixed in the killing gauge \cite{{Pesando},{KalloshRahmfeld}}, $\Theta^1=\Gamma_{0123} \Theta^2$, where $\Theta^1$ and $\Theta^2$ are 32-component Majorana-Weyl spinors, and $\Gamma_{0123}$ is an antisymmetric combination of the $SO(9,1)$ gamma matrices $\Gamma^M$. The partition function is given by 
\beqa
Z_{\mbox{full}}
&=&
\sum_{h} e^{\gamma(2h-2)} \int\mathcal{D}g\int\mathcal{D}x\int\mathcal{D}u\int\mathcal{D}\hat{y}\int\mathcal{D}\psi  \n
&&
\exp\left[
\frac{1}{\sqrt{\lambda}}
\int d^2 \sigma 
\left( 
\frac{1}{2}\sqrt{-g}
\left(
\frac{(\partial_{\alpha} x^{\mu}-2i\psi^T\gamma^{\mu} \partial_{\alpha}\psi)^2  + \lambda (\partial_{\alpha} u)^2 )}{u^2}
+ \lambda(\partial_{\alpha} \hat{y}^I)^2 
\right) \right. \right. \n
&& 
\qquad \qquad \qquad \qquad -2i \left. \left. \sqrt{\lambda} \epsilon^{\alpha \beta} \frac{\partial_{\alpha}\psi^T\hat{y}^I\gamma^I \partial_{\beta} \psi}
{u}
\right)\right],
\eeqa
where $\sigma^{\alpha}$ $(\alpha=1,2)$ represents the world-sheet coordinates, $e^{\gamma}$ is the string coupling constant, and $h$ is the number of genuses. Also, $x^{\mu}$, $u$, $\hat{y}^I (\hat{y}^2=1)$ and $\psi$ represent the four-dimensional scalars, the radial coordinate, the six spherical coordinates of $S^5$ and the 16 component Majorana-Weyl fermions, respectively. The quantities $\gamma^{\mu}$ and $\gamma^{I}$ are defined in (\ref{gamma}). We choose this gauge because we have four-dimensional Lorentz covariance and supersymmetry in it. 

By discretizing the world-sheets and taking the $\lambda \to 0$ limit of the action defined on each discretized world-sheet, we obtain the sum of the leading terms\footnote{$\int \mathcal{D}\hat{y}1=1$ is used.},
\beqa
Z
&=&
\sum_{h} e^{\gamma(2h-2)} \sum_{\mbox{random lattices}}  \int\mathcal{D}x\int\mathcal{D}u\int\mathcal{D}\psi \exp\left(-\frac{1}{2\sqrt{\lambda}} \sum_{\langle i,j\rangle  }\frac{(x_i^{\mu}-x_j^{\mu}+2i\psi_i^T\gamma^{\mu}\psi_j)^2}{u_iu_j}\right) \n
&=& \sum_{h} e^{\gamma(2h-2)} \sum_{\mbox{random lattices}}
Z_{\mbox{random}}, \label{pathintegral}
\eeqa
where $\sum_{\mbox{random lattices}}$ represents the sum over arbitrary discretized world-sheets that are constructed from triangles and quadrangles. We call $Z_{\mbox{random}}$ the partition function for a discretized world-sheet. This limiting theory is equivalent to the theory that is obtained by taking the geometric zero radius limit of the $AdS_5 \times S^5$ superstring with the kappa symmetry fixed in the killing gauge introduced in \cite{NastaseSiegel}, where a new AdS/CFT correspondence and a Feynman rule of $\mathcal{N}=4$ SYM in a superspace are proposed.\footnote{In contrast, in our paper, we investigate this theory in the context of the original AdS/CFT correspondence proposed by Maldacena.} By rescaling the fields and removing $\lambda$ in the action of (\ref{pathintegral}), we find that the path-integral measures produce a factor\footnote
{Unlike S-matrices, unrenormalized partition functions are not invariant under such a rescaling.} 
 of $\lambda$, and we obtain 
\beq
Z=\sum_{h} e^{\gamma(2h-2)} \sum_{\mbox{random lattices}} \lambda^{n_r} \int\mathcal{D}x\int\mathcal{D}u\int\mathcal{D}\psi \exp\left(-\frac{1}{2} \sum_{\langle i,j\rangle  }\frac{(x_i^{\mu}-x_j^{\mu}+2i\psi_i^T\gamma^{\mu}\psi_j)^2}{u_iu_j}\right), \label{pathintegral2}
\eeq
where $n_r$ is an integer-valued function of the number of faces, edges and vertices in the triangles and quadrangles on the random lattice.\footnote
{In the $\lambda \to 0$ limit, the partition function (\ref{pathintegral2}) vanishes. This behavior is consistent with the fact that the SYM becomes a free theory in this limit.}
We cannot determine $n_r$ because of the ambiguity inherent in the definition of the path integrals. For example, naive rescaling $u_i$ and rescaling $x^{\mu}_i$ and $\psi_i$ produce different factors. In general, it is difficult to determine the factor for a path integral. We plan to study this problem in the future. In order to make the quartic fermion term quadratic, we rewrite this path integral in its first-order form:
\beqa
Z&=&\sum_{h} e^{\gamma(2h-2)} \sum_{\mbox{random lattices}}  \lambda^{n_r} 
\int\mathcal{D}x\int\mathcal{D}u\int\mathcal{D}\psi \int \mathcal{D}p \n
&&
\exp\left(-\frac{1}{2} \sum_{\langle i,j\rangle  }\left(u_iu_j(p_{ij}^{\mu})^2+2ip_{ij}^{\mu}(x_i^{\mu}-x_j^{\mu}+2i\psi_i^T\gamma^{\mu}\psi_j)\right)\right).
\label{Z}
\eeqa
If the equation of motion for $p_{ij}^{\mu}$ is used, (\ref{Z}) reduces to (\ref{pathintegral2}).
We can easily integrate $x^{\mu}$, $u$ and $\psi$ in (\ref{Z}) and obtain the effective partition function for $p^{\mu}_{ij}$,
\beq
Z=\sum_{h} e^{\gamma(2h-2)} \sum_{\mbox{random lattices}}  \lambda^{n_r} 
\int \mathcal{D}p \: \mbox{det}^{-1/2}(p^2_{ij}) \:
\mbox{det}^{1/2}(p_{ij}^{\mu}\gamma_{\mu})
\prod_{i} \delta^4(\sum_{j} p^{\mu}_{ij}). \label{stringresult}
\eeq

Finally, we compare the formal free energy of the SYM given in (\ref{SYMresult}) and the formal partition function of the superstring given in (\ref{stringresult}). More precisely, we compare each bubble diagram in (\ref{SYMresult}) and the partition function for the corresponding discretized world-sheet $Z_{\mbox{random}}$ in (\ref{stringresult}).  The common properties of (\ref{SYMresult}) and (\ref{stringresult}) are as follows. First, the continuous degrees of freedom in (\ref{SYMresult}) and (\ref{stringresult}) are the same as those in $k_{ij}^{\mu}$ and in $p_{ij}^{\mu}$, respectively, which are defined at the nearest neighbors on the random surfaces. This result implies that the Fourier momenta $k_{ij}^{\mu}$ of the four-dimensional coordinates in the SYM should be identified with the canonical momenta $p_{ij}^{\mu}$ of the four-scalars in the string. Second, (\ref{SYMresult}) and (\ref{stringresult}) depend on $k_{ij}^{\mu} (=p_{ij}^{\mu})$ only through $k^2_{ij}$ and $k_{ij}^{\mu}\gamma_{\mu}$, because they are Lorentz invariant in four-dimensional spaces. Third, on every corresponding cite, (\ref{SYMresult}) and (\ref{stringresult}) have the same delta function, which represents momentum conservation on each cite in the SYM. In order to compare (\ref{SYMresult}) and (\ref{stringresult}) more precisely, we need to perform the summation over $M_{ij}$ in (\ref{SYMresult}). 


\vspace*{1cm}

\section{Summary and Discussion}
\setcounter{equation}{0}

We obtained the formal free energy of $\mathcal{N}=4$ $U(N)$ super Yang-Mills theory in four dimensions generated by the Feynman graph expansion to all orders of the weak 't Hooft coupling with arbitrary $N$. This free energy is written as the sum over the discretized, closed two-dimensional surfaces that are identified with the world-sheets of the dual string. We compared, to all orders of the string coupling constant, this free energy with the formal partition function of the discretized $AdS_5 \times S^5$ superstring with the kappa-symmetry fixed in the killing gauge and in the expansion corresponding to the weak 't Hooft coupling expansion in the SYM. We found certain properties common to this free energy and partition function. This supports a conjecture that the free energy of the $\mathcal{N}=4$ SYM corresponds to the partition function of the $AdS_5 \times S^5$ superstring even if the corresponding operators are inserted into both theories. The formal partition function and the formal free energy have non-trivial forms and seem to correspond to each other. Moreover, even if we insert operators that depend only on $p_{ij}^{\mu}$ ($\Omega(p)$) into the formal partition function of the string theory and the same operators ($\Omega(k)$) into the formal free energy of the SYM, the above correspondence still holds. It would be interesting to examine whether these operators are consistent with the known corresponding operators in the AdS/CFT correspondence. 

We cannot insert operators that consist of the radial coordinate and the $S^5$ coordinates into the partition function of the superstring at leading order in $\lambda$. This implies that we can treat only the four-dimensional sector of the superstring in our perturbative treatment. It is important to study whether the free energy of the SYM corresponds to the partition function of the superstring when the full corresponding operators are inserted. 

Further studies\footnote
{There is a possibility that the formal free energy of the SYM and the formal partition function of the string have slightly different forms and that they belong to the same universality class, even if the free energy of the SYM corresponds to the partition function of the string. This is the limitation of our method, as one can see the difference between ($\ref{c=1 matrix}$) and ($\ref{c=1disc}$) in the \textit{c}=1 case. 
}
 are required to obtain a more precise comparison. First, we need to calculate the sum over $M_{ij}$ in the free energy (\ref{SYMresult}). Second, we need to determine the dependence on $\lambda$ in the partition function (\ref{stringresult}). Third, we need to calculate the corrections of higher order in $\lambda$ to the partition function (\ref{stringresult}). 

Our studies show that for arbitrary $\lambda$, individual bubble diagrams of the SYM depend only on $\lambda^V$, whereas individual partition functions for discretized world-sheets $Z_{\mbox{random}}$ have infinitely many forms of $\lambda$ dependence, which come from the higher-order corrections in general. If there is no higher-order correction, because of supersymmetry, we expect that $Z_{\mbox{random}}$ depends only on $\lambda^{V}$ on the string side and that each bubble diagram of the SYM corresponds to $Z_{\mbox{random}}$. Even if there are higher-order corrections, we expect that in total, the sum of all bubble diagrams of the SYM corresponds to the sum of the partition functions for all discretized world-sheets of the superstring in the continuum limit, because the functional forms in  (\ref{SYMresult}) and (\ref{stringresult}) do not depend on the configurations of random lattices. 

We should comment on a continuum limit of the $\mathcal{N}=4$ SYM
\footnote{This continuum limit is different from the continuum limit in the SYM on a four-dimensional lattice.}
in which world-sheets become continuous. The free energy of the $\mathcal{N}=4$ SYM, which is non-perturbatively defined in the 't Hooft coupling $\lambda$, is a regular real function, because the $\mathcal{N}=4$ SYM is a conformal field theory for arbitrary $\lambda$. However, if $\lambda$ is defined on a complex plane, the free energy should have a pole at a complex value of $\lambda$, because the free energy depends on $\lambda$. Therefore, the $\lambda$ expansion of the free energy has a radius of converge $\lambda_c$. The value of $\lambda_c$ is given by the distance between the origin and the pole. Actually, $\lambda$ expansions of correlation functions are expected to have finite radii of convergence in Yang-Mills theories \cite{finite}. If $\lambda$ approaches $\lambda_c$ (the continuum limit), the $\lambda$ expansion diverges. This divergence implies that terms with infinitely many interactions become dominant in the $\lambda$ expansion and that world-sheets become continuous as in the \textit{c}=1 matrix model case. In our analysis, $\lambda$ in the SYM behaves like $g$ in the \textit{c}=1 matrix model. Therefore, the renormalized difference between $\lambda$ and $\lambda_c$ should be $\frac{R^4}{\alpha'^2}$ in the superstring, analogous to the cosmological constant in the \textit{c}=1 string theory.

In summary, our result suggests a mechanism for the dynamic emergence of the world-sheet from $\mathcal{N}=4$ SYM. Therefore it enables us to determine how the $AdS_5 \times S^5$ superstring is reproduced in the AdS/CFT correspondence.

\vspace*{1cm}

\section*{Acknowledgements}
I would like to thank H. Kawai and Y. Kitazawa for stimulating discussions. I am also grateful to A. Tsuchiya not only for stimulating discussions but also for useful comments on the manuscript. This work is supported in part by the U.S. DOE Grant No. DE-FG02-91ER40685.

\vspace*{1cm}

\section*{Appendix A: Feynman rules}
\setcounter{equation}{0}
\renewcommand{\theequation}{A.\arabic{equation}}
In this appendix, we summarize the Feynman rules for $\mathcal{N}=4$ $U(N)$ super Yang-Mills theory in the double-line notation.

\noindent
\underline{Propagators}

\noindent
We formally introduce $\hat{\psi}$ to count the minus signs resulting from fermion interchanges, as in (\ref{modification}). All propagators among $\psi$ and $\hat{\psi}$ have the same values. The propagators are given by 
\beqa
\langle A^{\mu}_{ij}(x)A^{\nu}_{kl}(y)\rangle  &=&\eta^{\mu\nu}\delta_{il}\delta_{jk} \Delta(x-y)\n
\langle X^{I}_{ij}(x)X^{J}_{kl}(y)\rangle  &=&\delta^{IJ}\delta_{il}\delta_{jk} \Delta(x-y)\n
\langle c_{ij}(x) \bar{c}_{kl}(y)\rangle  &=&\delta_{il}\delta_{jk} \Delta(x-y)\n
\langle \psi_{ij}(x)\psi_{kl}(y)\rangle  &=&\langle \hat{\psi}_{ij}(x)\psi_{kl}(y)\rangle  =\langle \psi_{ij}(x)\hat{\psi}_{kl}(y)\rangle  =\langle \hat{\psi}_{ij}(x)\hat{\psi}_{kl}(y)\rangle  =\delta_{il}\delta_{jk} \Delta_s(x-y), \n
\label{SYMpropagator}
\eeqa
where
\beq
\Delta(x-y) = g_{YM}^2 \int d^4k \frac{-i}{k^2} e^{ik(x-y)},
\label{bosonicpropagator}
\eeq
\beq
\Delta_s (x-y) = g_{YM}^2 \int d^4k \frac{-ik_{\mu}\hat{\gamma}^{\mu}}{k^2} e^{ik(x-y)}.
\label{fermionicpropagator}
\eeq

\noindent
\underline{Vertices}

\begin{figure}[htbp]
\begin{center}
\psfrag{mixed}
{$
\frac{i}{g_{YM}^2}(2\eta_{M_1M_3}\eta_{M_2M_4}-\eta_{M_1M_2}\eta_{M_3M_4}-\eta_{M_1M_4}\eta_{M_2M_3})
$}
\psfrag{phi}{$\phi^M=(A^{\mu}, X^{I})$}
\psfrag{M1}{$\phi^{M_1}$}
\psfrag{M2}{$\phi^{M_2}$}
\psfrag{M3}{$\phi^{M_3}$}
\psfrag{M4}{$\phi^{M_4}$}
\includegraphics[height=3cm, keepaspectratio, clip]{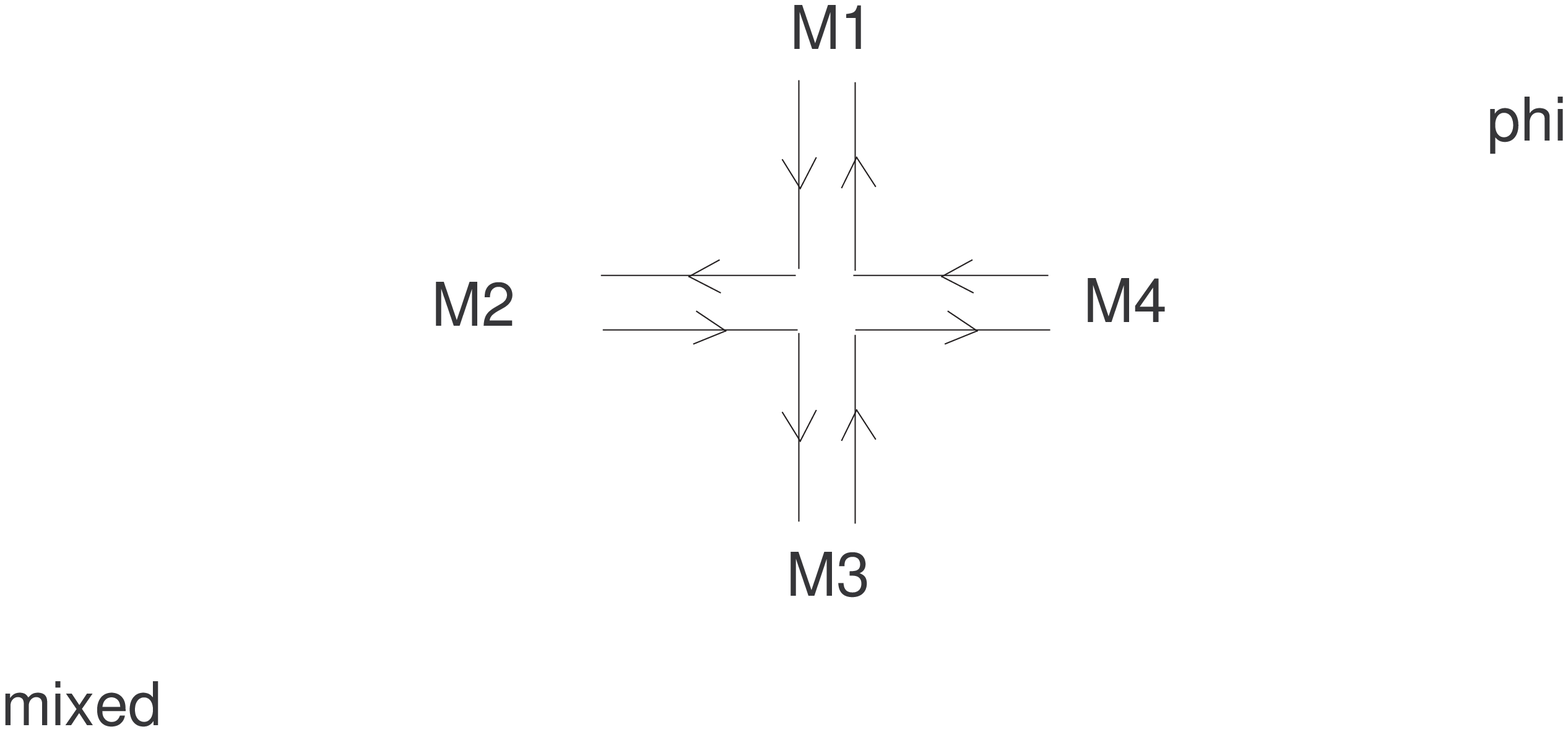}\end{center}
\caption{Four-point interactions.}
\label{4ptfeynman}
\end{figure}

\begin{figure}[htbp]
\begin{center}
\psfrag{k1}{$k_1$}
\psfrag{k2}{$k_2$}
\psfrag{k3}{$k_3$}
\psfrag{a0}{$A^{\mu}$}
\psfrag{a1}{$A^{\mu_1}$}
\psfrag{a2}{$A^{\mu_2}$}
\psfrag{a3}{$A^{\mu_3}$}
\psfrag{x0}{$X^{I}$}
\psfrag{x1}{$X^{I_1}$}
\psfrag{x2}{$X^{I_2}$}
\psfrag{x3}{$X^{I_3}$}
\psfrag{c}{$c$}
\psfrag{c'}{$\bar{c}$}
\psfrag{f}{$\psi$}
\psfrag{f'}{$\hat{\psi}$}
\psfrag{AAAinteractions}
{$
\frac{i}{g_{YM}^2}
\delta(k_1+k_2+k_3)
\left(
\eta_{\mu_1\mu_2}(k_1^{\mu_3}-k_2^{\mu_3})
+(\mbox{cyclic in (1,2,3)})
\right)
$}
\psfrag{AXXinteractions}
{$
\frac{i}{g_{YM}^2}
\delta(k_1+k_2+k_3)
\delta_{I_2I_3}(k_2^{\mu_1}-k_3^{\mu_1})
$}
\psfrag{ACC'interactions}
{$
\frac{i}{g_{YM}^2}
\delta(k_1+k_2+k_3)
k_2^{\mu}
$}
\psfrag{AC'Cinteractions}
{$
-\frac{i}{g_{YM}^2}
\delta(k_1+k_2+k_3)
k_3^{\mu}
$}
\psfrag{AFFinteractions}
{$
-\frac{i}{g_{YM}^2}
\delta(k_1+k_2+k_3)
\gamma^{\mu}
$}
\psfrag{XFFinteractions}
{$
-\frac{i}{g_{YM}^2}
\delta(k_1+k_2+k_3)
\gamma^{I}
$}
\includegraphics[height=10cm, keepaspectratio, clip]{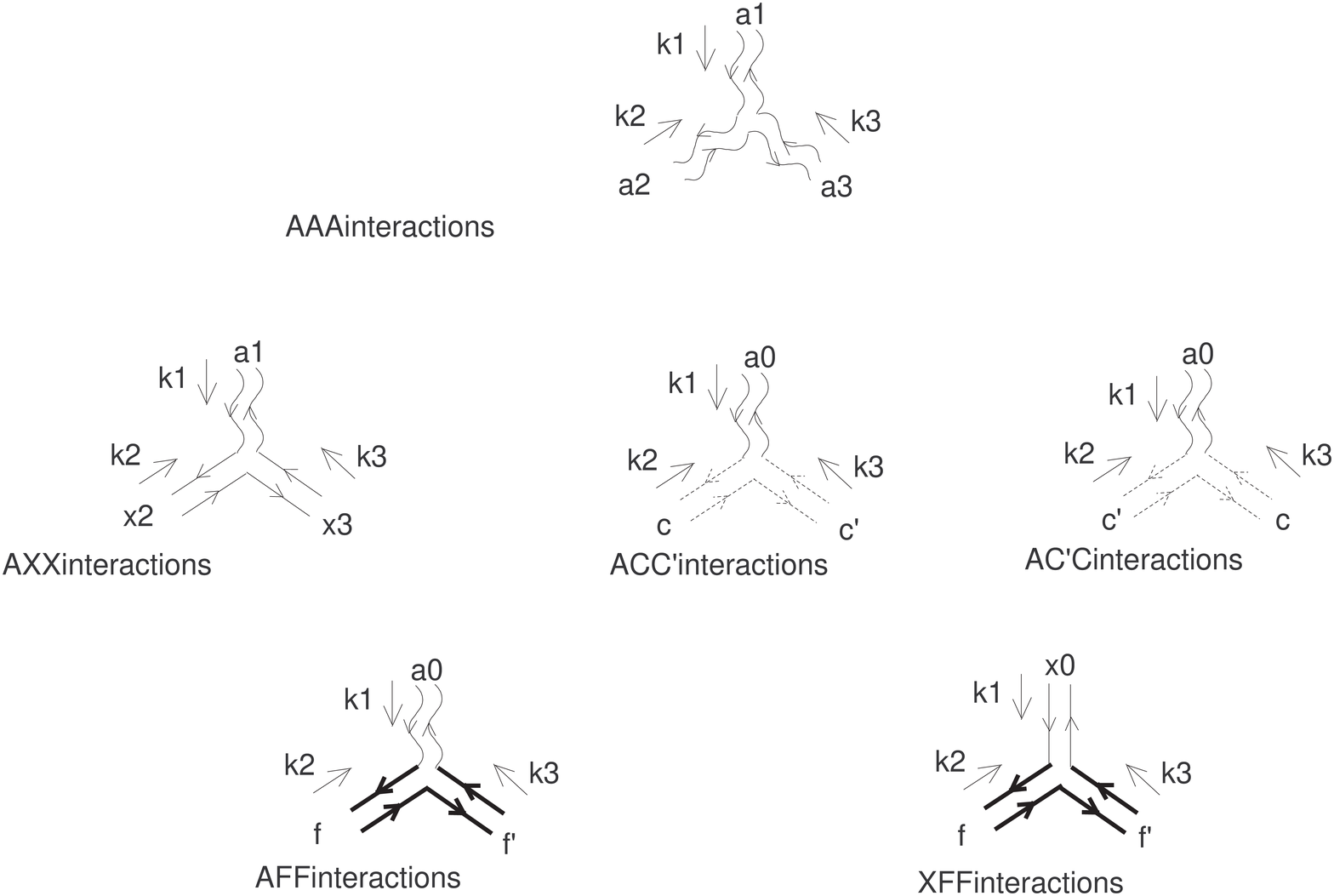}\end{center}
\caption{Three-point interactions.}
\label{3ptfeynman}
\end{figure}

\noindent
All the four-point interactions can be summarized with respect to one kind of vertex in Fig. \ref{4ptfeynman}, where 
$\phi^M=(A^{\mu}, X^{I})$, because $\mathcal{N}=4$ SYM in four dimensions can be obtained from $\mathcal{N}=1$ SYM in ten dimensions through dimensional reduction \cite{{BrinkSchwarzScherk},{GliozziScherkOlive}}. The three-point interactions are summarized in Fig. \ref{3ptfeynman}.
%
%
%
%

\vspace*{1cm}

\section*{Appendix B: Formal Free Energy of SYM in Ten Dimensions}
\setcounter{equation}{0}
\renewcommand{\theequation}{B.\arabic{equation}}
We can easily obtain the formal free energy of $\mathcal{N}=1$ SYM in ten dimensions in the same way as that of $\mathcal{N}=4$ SYM in four dimensions:
\beqa
F_{\mbox{10 dim.}}
&=&
\sum_{\mbox{random lattices}}\sum_{M_{ij}=0, \cdots, 15}N^{2-2h}\lambda^{E-F}(-1)^{\mid\mbox{fermion loops}\mid+\mid\mbox{fermion interchanges}\mid} \n
&& \!\!\!\!
\prod_{\langle i,j\rangle  (i<j)} 
\left[
\int d^{12} k_{ij} \delta^2(k^a_{ij})
\left(
\sum_{s=12}^{15}(1-\delta_{M_{ij},s})\Delta_{10}(k_{ij}^{M})
+\sum_{s=12}^{15}\delta_{M_{ij},s}\Delta_{10 s}^{ij}(k_{ij}^{M})
\right)
\right] \n 
&& \!\!\!\!
\prod_{i(4pt.)}
 \left[
\delta^{10}(\sum_{p=1}^4k_{ij_p})
\prod_{t=10}^{15}\prod_{e=1}^4(1-\delta_{M_{ij_e},t}) 
(2\eta_{M_1M_3}\eta_{M_2M_4}-\eta_{M_1M_2}\eta_{M_3M_4}-\eta_{M_1M_4}\eta_{M_2M_3})
\right]
\n
&& \!\!\!\! \prod_{i(3pt.)}
\delta^{10}(\sum_{q=1}^3k_{ij_q})
\Biggl[
\prod_{u=12}^{15}\prod_{f=1}^3(1-\delta_{M_{ij_f},u})  
\Biggl( \eta_{M_{ij_1}M_{j_2i}}\Bigl((1-\delta_{M_{ij_1}, 10})k_{ij_1}^{M_{ij_3}}-(1-\delta_{M_{ij_2}, 10})k_{ij_2}^{M_{ij_3}}\Bigr)\n&& \qquad\qquad\qquad\qquad\qquad\qquad\qquad\qquad\qquad\qquad\qquad +\left(\mbox{cyclic in (1,2,3)}\right) \Biggr) \n
&& \qquad\qquad\qquad\quad
+\Bigl(
(-1)^{1+\Theta(i-j_2)+\Theta(i-j_3)}
(\delta_{M_{ij_2},12}+\delta_{M_{ij_2},13})
(\delta_{M_{ij_3},14}+\delta_{M_{ij_3},15})
\gamma_i^{M_{ij_1}} \n
&& \qquad\qquad\qquad\qquad\qquad\qquad\qquad\qquad\qquad\qquad\qquad
+\left(\mbox{cyclic in (1,2,3)}\right)
\Bigr)
\Biggr]. \nonumber
\eeqa
Here, we have
\beq
\Delta_{10} (k^{M}) = g_{YM}^2  \frac{-i}{k^2},
\eeq
\beq
\Delta_{10 s} (k^{M}) = g_{YM}^2 \frac{-ik_{M}\hat{\gamma}^{M}}{k^2},
\eeq
where $M$ runs from 0 to 9.

\end{document}